\documentclass[aps,twocolumn,showpacs,groupedaddress,letter]{revtex4-1}
\usepackage{graphicx,float}
\usepackage{caption}
\usepackage{amssymb,amsmath,bm}
\usepackage{color}
\begin{document}

\title{Physical Principles of the Amplification of Electromagnetic Radiation
Due to Negative Electron Masses in a Semiconductor Superlattice}

\author{A.\,V.\,Shorokhov}
\affiliation{Mordovian National Research University, Saransk 430005, Russian Federation}
\email[]{alex.shorokhov@mail.ru}
\author{M.\,A.\,Pyataev}
\affiliation{Mordovian National Research University, Saransk 430005, Russian Federation}
\author{N.\,N.\,Khvastunov}
\affiliation{Mordovian State Pedagogical Institute, Saransk 430007, Russian Federation}
\author{T.\,Hyart}
\affiliation{Department of Physics, University of Jyv\"{a}skyl\"{a},
Jyv\"{a}skyl\"{a} FI-40014, Finland}
\author{F.\,V.\,Kusmartsev}
\affiliation{Department of Physics, Loughborough University, Loughborough LE11
3TU, United Kingdom}
\author{K.\,N.\,Alekseev}
\affiliation{Department of Physics, Loughborough University, Loughborough LE11
3TU, United Kingdom}


\begin{abstract}
In a superlattice placed in crossed electric and magnetic
fields, under certain conditions, the inversion of electron
population can appear at which the average energy of electrons is
above the middle of the miniband and the effective mass of the
electron is negative. This is the implementation of the negative
effective mass amplifier and generator (NEMAG) in the superlattice.
It can result in the amplification and generation of terahertz
radiation even in the absence of negative differential conductivity.
\end{abstract}
\pacs{67.57.Lm, 76.60.-k}
\maketitle

The idea of using the effect of negative effective mass of charge
carriers in semiconductors for creating negative effective mass
amplifier and generator (NEMAG) was proposed by H. Kr\"{o}mer
\cite{Kromer58} in 1958 by an example of holes with a negative
effective mass in germanium placed in an electric field. It was
experimentally implemented for the first time in the mid-1980s
\cite{Andronov84}. The experimental and theoretical works in this
field were reviewed in \cite{Andronov87}. Conditions for the
realization of the NEMAG become particularly favorable when a
bulk semiconductor is additionally placed in a static magnetic field
\cite{Andronov86}.

In this work, we show that NEMAG can also be implemented in a
semiconductor superlattice operating in a miniband transport
regime when it is placed in crossed electric and magnetic fields
of moderate strengths. In this case, under certain conditions, a
population inversion can arise at which the average energy of
electrons is above the middle of the miniband and the effective
electron mass becomes negative in average. This in turn
results in the amplification of terahertz radiation in the absence
of instabilities caused by negative differential conductivity and
being characteristic of the classical amplification schemes
\cite{Esaki70,Ktitorov72}. In contrast to quantum cascade
structures of weakly coupled wells, where the achievement of an
inverse population between Landau levels allows the creation of a
tunable quantum generator of terahertz radiation \cite{Jasnot12},
the dynamics of electrons in the strongly coupled superlattice is
in essence semiclassical \cite{Shik73,Bauer02}. Note that
terahertz electroluminescence in the SiC structure having a
natural superlattice was recently observed in \cite{Sankin11}.
This indicates the possibility of generating terahertz radiation
in superlattice structures operating in miniband transport regime.

The minimal semiclassical model of the superlattice in
crossed fields was proposed by Polyanovskii \cite{Pol80}. He
analytically found the current and its dependence on the field
strengths. As follows from the results reported in
\cite{Sibille90,Palmier92}, this simple single miniband model
satisfactorily describes the experimental current–voltage
characteristics. Various electronic and optical properties of
superlattices, demonstrating the single miniband transport regime
under the action of electric and magnetic fields of various
configurations, were studied in \cite{Bass80}-\cite{Selskii11}.
Finally, the significant wideband amplification of terahertz waves
in a superlattice placed in crossed fields was recently predicted
in \cite{Hyart09} on the basis of numerical analysis.

However, the physical principles of amplification in such a system
can hardly be revealed by numerical analysis alone. The
understanding of these principles is not only of fundamental
interest but also necessary for the choice of the optimal
parameters of an amplifier. In this work, we combine the
analytical \cite{Pol80} and numerical \cite{Hyart09} methods to
demonstrate the feasibility of NEMAG in the
superlattice and to clarify its relation to the previously
described terahertz amplification regime.

The analysis of electron trajectories in the phase space makes it
possible to visually understand the physical essence of the
effect. For this reason, we first consider the ballistic transport
regime for electrons in the superlattice placed in a static
electric field $\bf{E}$, which is directed along the axis of the
superlattice ($x$ axis), and a static magnetic field $\bf{H}$,
which is directed along the $z$ axis. In the tight-binding
approximation, the dependence of the energy $\varepsilon$ on the
quasimomentum ${\bf p}$ has the standard form
\begin{equation}
\label{energy}
\varepsilon({\bf p})=\frac{\Delta}{2}\left(1-\cos\frac{p_xd}{\hbar}\right)+\frac{1}{2m_\perp}(p_y^2+p_z^2),
\end{equation}
where $\Delta$ is the width of the miniband, $d$ is the period of
the superlattice, and $m_\perp$ is the effective mass of the
electron along the layers of the superlattice. Using the equation
of motion
\begin{equation}
\mathbf{\dot{p}}=e\mathbf{E}+\frac{e}{c}\mathbf{V}\times\mathbf{H}
\end{equation}
and the standard formula for the electron velocity, ${\bf
V}=\partial\varepsilon({\bf p})/\partial {\bf p}$, we obtain the
following semiclassical equations describing the ballistic
transport regime:
\begin{equation}
  \label{systNewtonx}
  \left\{\begin{array}{l}
    \dot{K}_x=\omega_B+\omega_cK_{y},\\
    \dot{K}_y=-\omega_c\sin{K_x}.
    \end{array}\right.
\end{equation}
Here, $K_x=p_xd/\hbar$, $K_y=\sqrt{m_0/m_\perp}p_yd/\hbar$,
$\omega_B=eEd/\hbar$ is the Bloch frequency, $m_0=2\hbar^2/\Delta
d^2$ is the effective electron mass at the bottom of the
miniband along the axis of the superlattice, and
$\omega_c=eH/\sqrt{m_xm_\perp}c$ is the effective cyclotron
frequency. The momentum along the $z$ axis is conserved.

According to Eq. (\ref{systNewtonx}), the projection of the
quasimomentum on the $x$ axis, which governs the motion of
electrons along the superlattice axis, satisfies the
pendulum equation
\begin{eqnarray}
\label{pend}
\ddot{K}_x+\omega_c^2\sin K_x=0.
\end{eqnarray}
We note that the electric field enters into this equation only
through the initial conditions. The amplification of external radiation in this
system is possible only in the nonlinear regime, because this
equation at small $K_x$ values is a usual equation of small
oscillations that allows only resonant (cyclotron) absorption of
radiation, rather than amplification.

From Eq. (\ref{pend}) we obtain the following equation of a phase
trajectory in the quasimomentum space:
\begin{eqnarray}
\label{genint}
\left(\dfrac{\omega_B}{\omega_c}
+K_y\right)^2
+4\sin^2{\left(\dfrac{K_x}{2}\right)}=
\left(\dfrac{\Omega}{\omega_c}\right)^2,
\end{eqnarray}
where $\Omega$ is determined by the initial conditions for the
quasimomentum, $K_x(0)=K_x^0$ and $K_y(0)=K_y^0$, and by the
electric and magnetic field strengths.

In the general case, the frequency of nonlinear oscillations has
the form \cite{Pol80}
\begin{eqnarray}
\Omega_{eff}=
\left\{\begin{array}{l}
    \dfrac{\pi\omega_c}{2\mathbf{K}(\Omega/2\omega_c)},  \Omega<2\omega_c\\
    \\
    \dfrac{\pi\Omega}{2\mathbf{K}(2\omega_c/\Omega)},  \Omega>2\omega_c,
    \end{array}\right.
\end{eqnarray}
where ${\bf K}(x)$ is the complete elliptic integral of the first
kind. We note that such oscillations have been observed experimentally
\cite{Bauer02}.

We now indicate some characteristic features of ballistic
trajectories of electrons, which can be conveniently analyzed by
considering trajectories in the quasimomentum space (Fig. 1).

\begin{figure}[h]
\centering
\includegraphics[clip=true,width=0.5\textwidth]{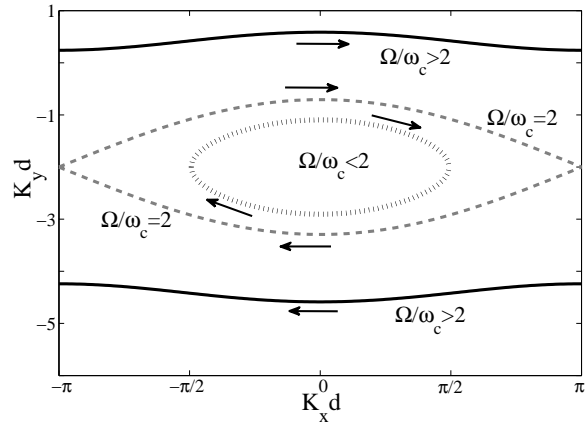}%
\caption{Ballistic trajectories in the $(K_x; K_y)$ plane.}
\label{fig:1}
\end{figure}

The equation of the pendulum (\ref{pend}) has a separatrix
$2\omega_c=\Omega$ separating physically different regimes of the
periodic motion of electrons. No oscillations exist on the
separatrix. Here electrons only asymptotically approach the top
of the miniband. However, this trajectory is unstable against any
noise or fluctuations. For $2\omega_c>\Omega$, the magnetic field
is strong enough to confine electrons in the first Brillouin zone,
and quasi-cyclotron oscillations occur. With an increase in the
magnetic field, trajectories become more similar to usual cyclotron
trajectories. In this case, electrons oscillate near the bottom of
the miniband. For $\Omega/\omega_c>2$, electrons undergo the
complete motion over the miniband, which is accompanied by Bragg
reflections. The frequency of Bloch oscillations is modified owing
to the magnetic-field-induced bend of trajectories. It is
noteworthy that the frequencies of both quasi-Bloch and
quasi-cyclotron oscillations decrease upon approaching the
separatrix. The choice of a certain type of motion of electrons
strongly depends on their initial quasimomentum. In a particular
case of zero initial conditions ($K_x(0)=0, K_y(0)=0$) the
equation of the pendulum has the separatrix at
$2\omega_c=\omega_B$. These zero initial conditions correspond to
the case of low temperatures and low electron densities, where
only states near the bottom of the miniband are occupied in
equilibrium. This case will be considered in what follows.

If the scattering of electrons is taken into account within the
approximation of constant relaxation time $\tau$, the separatrix
at low temperatures separates the regions of the positive and
negative slopes in the current–voltage characteristic of the
superlattice \cite{Pol80,Hyart09} (with allowance for the shift of
the maximum by a value of about $\tau^{-1}$, characteristic of
dissipative systems). This is quite clear because the electron
undergoing quasi-cyclotron oscillations does not reach the edge of
the Brillouin zone and does not undergo total reflection. Thus,
the generation of high-frequency radiation without negative
differential conductivity is in principle possible only for
quasi-cyclotron oscillations, which correspond to the oscillational
rather than rotational regime of pendulum (\ref{pend}).

\begin{figure}[h]
\centering
\includegraphics[clip=true,width=0.5\textwidth]{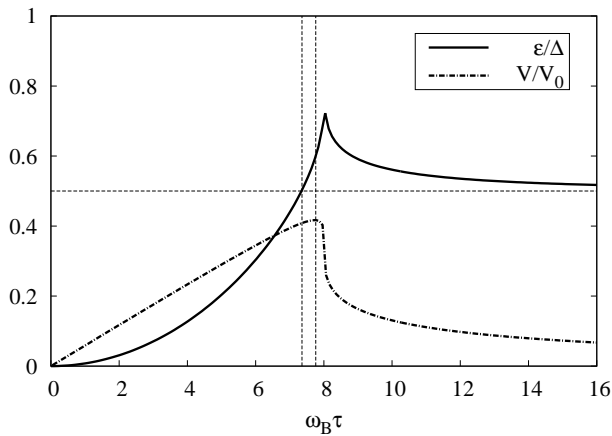}%
\caption{Average energy (solid line) and drift velocity (dash-dotted
line) versus the scaled electric field $\omega_B\tau$ at
$\omega_c\tau=4$. The velocity is normalized to the maximum miniband
velocity of the electron $V_0=\Delta d/2\hbar$ and the energy is
normalized to $\Delta$. The horizontal dashed straight line
corresponds to the middle of the miniband and the vertical dashed
straight lines correspond to Bloch frequencies at which the
effective mass and derivative of the velocity with respect to
$\omega_B$ change sign.} \label{fig:2}
\end{figure}

In physical terms, the possibility of the amplification of
radiation is due to a specific behavior of the average energy and
effective mass. We calculate the average energy of electrons along
the $x$ axis under the assumption that they are on the bottom of
the miniband at the initial time:
\begin{eqnarray}
\left\langle \varepsilon_x \right\rangle=\dfrac{1}{\tau}\int_{0}^{\infty}{\varepsilon_x(t) {\rm{e}}^{-t/\tau}dt}.
\end{eqnarray}
Here,
\begin{eqnarray}
\varepsilon_x=\frac{\Delta}{2}\left[1-\cos\frac{p_xd}{\hbar}\right]=\Delta\ {\rm sn}^2\left(\frac{\omega_B t}{2}; \frac{2\omega_c}{\omega_B}\right),
\end{eqnarray}
where ${\rm sn}(u;k)$ is the elliptic Jacobi function.

Then, at $\omega_B/\omega_c>2$, we obtain
\begin{eqnarray}
\label{energy1} \left\langle \varepsilon_x
\right\rangle&=&\frac{2\pi^2\Delta}{k^2{\bf K}^2(k)}
\sum_{n,l=0}^{\infty}
\frac{q^{n+l+1}}{\left(1-q^{2n+1}\right)\left(1-q^{2l+1}\right)}\nonumber\\
&\times&\left[\frac{1}{1+\left[(n-l)\pi\omega_B\tau/2{\bf K}(k)\right]^2}- \right.\nonumber\\
&-&\left.\frac{1}{1+\left[(n+l+1)\pi\omega_B\tau/2{\bf
K}(k)\right]^2}\right],
\end{eqnarray}
Here, $k=2\omega_c/\omega_B$ and $q=\rm{e}^{-\pi\bf{K'}/\bf{K}}$,
where $\bf{K'}$ is the complete elliptic integral of the first
kind of the argument $\sqrt{1-k^2}$. According to Eq.
(\ref{energy1}) and Fig. 2, the average energy in the limit
$\omega_B\rightarrow\infty$ tends to the middle of the band
($\left\langle \varepsilon_x \right\rangle\rightarrow\Delta/2$),
which is typical of conventional Bloch oscillations.

The average energy at $\omega_B/\omega_c<2$ has the form
\begin{eqnarray}
\label{energy2} \left\langle
\varepsilon_x\right\rangle&=&\frac{2\pi^2\Delta}{{\bf
K}^2}\sum_{n,l=0}^{\infty}\frac{q^{n+l+1}}{\left(1-q^{2n+1}\right)\left(1-q^{2l+1}\right)}\nonumber\\
&\times&\left[\frac{1}{1+\left[(n-l)\pi\omega_c\tau/{\bf K}\right]^2} \right.\nonumber\\
&-&\left.\frac{1}{1+\left[(n+l+1)\pi\omega_c\tau/{\bf K}\right]^2}\right],
\end{eqnarray}
where $k=\omega_B/2\omega_c$ in contrast to Eq. (\ref{energy1}).

At $\omega_B=2\omega_c$, we obtain
\begin{eqnarray}
\label{energy3} \left\langle \varepsilon_x
\right\rangle&=&-3+\dfrac{2}{\omega_B\tau}\left[\rm{\psi}\left(\dfrac{\omega_B\tau+1}{2\omega_B\tau}\right)
-\rm{\psi}\left(\dfrac{1}{2\omega_B\tau}\right)\right]\nonumber\\
&+&4\sum_{n,s=1}^{+\infty}{\dfrac{(-1)^{n+s}}{1+(n+s)\omega_B\tau}},
\end{eqnarray}
where ${\rm \psi}(x)$ is the Euler psi function.

The possibility of amplifying high-frequency radiation in the
superlattice without magnetic field is traditionally attributed to
the use of the negative differential conductivity regime, which
corresponds to the condition $\omega_B\tau>1$ \cite{Esaki70}.
The magnetic field not only shifts the maximum of the
current–voltage characteristic of the superlattice toward stronger
fields \cite{Sibille90} but also leads to the appearance of a new
region of amplification on the left of the peak of the
current–voltage characteristic, as well as to a significant
increase in the magnitude of amplification at negative
differential conductivity \cite{Hyart09}. We believe that these
amplification effects can be explained within the concept of
negative effective mass. The strongest amplification should be
expected when the average energy of electrons is above the middle
of the miniband and the effective mass becomes negative, as in the
NEMAG case, because according to the formula \cite{Ignatov93}
\begin{eqnarray}
m_x(\varepsilon_x)=\frac {m_0}{1-2\varepsilon_x/\Delta},
\end{eqnarray}
the transition to population inversion is directly related to the
appearance of negative effective mass.

Figure 2 shows the dependences of the average energy of electrons,
given by Eqs. (\ref{energy1}-\ref{energy3}), and their drift velocity
\cite{Pol80} on the normalized electric field $\omega_B\tau$. When
the electron gas is heated, electrons acquire energy. In conditions of
negative differential conductivity and at least for long superlattices
with ohmic contacts, this energy will be spent on the formation of propagating
high-field domains owing to the development of charge instabilities \cite{Alexeeva12}.
However, when the working point is chosen in the segment of the current-voltage
characteristic with a positive slope, instabilities are absent and
the collected energy can be directly transformed to radiation. The
maximum of the energy coincides with the point of transition from
rotational to oscillational regime ($\omega_B=2\omega_c$), whereas the
maximum of the average velocity is shifted leftward from this point
by a value of $\tau^{-1}$. Such a behavior of the energy and
velocity is well known in the theory of oscillations: Amplitude
resonance is always shifted from energy resonance by a value of
about the damping coefficient. It can be clearly seen in the figure
that the region where the average electron energy becomes above the
middle of the band begins with field values lower than the value
corresponding to the peak of the current–voltage characteristic.
Therefore, there exists a region near the separatrix where the average
energy of the electron is above the middle of the band, whereas the
slope of the current–voltage characteristic is positive and,
consequently, the system is stable. Such a system operates as a
classical analog of quantum generators of radiation.

The above statements are confirmed by the numerical analysis of
the absorption of the probe field $E(t)=E_\omega\cos(\omega t)$,
directed along the axis of the superlattice. The amplitude of the
probe field $E_\omega$ is small and the frequency $\omega$ varies
in a quite wide range. In this case, it is assumed that the value
of $\omega$ in experiments is governed by an external resonant
circuit. To calculate the absorption coefficient, we used the
path-integral method for solving the Boltzmann equation for the
superlattice \cite{Bass81,Hyart09}. For the numerical analysis, we
used the following typical parameters of the GaAs superlattice:
$d=6$~nm, $\Delta=60$~meV, density of free carriers
$n=10^{16}$~cm$^{-3}$, and dielectric constant $\varepsilon=13$.

\begin{figure}[h]
\centering
\includegraphics[clip=true,width=0.5\textwidth]{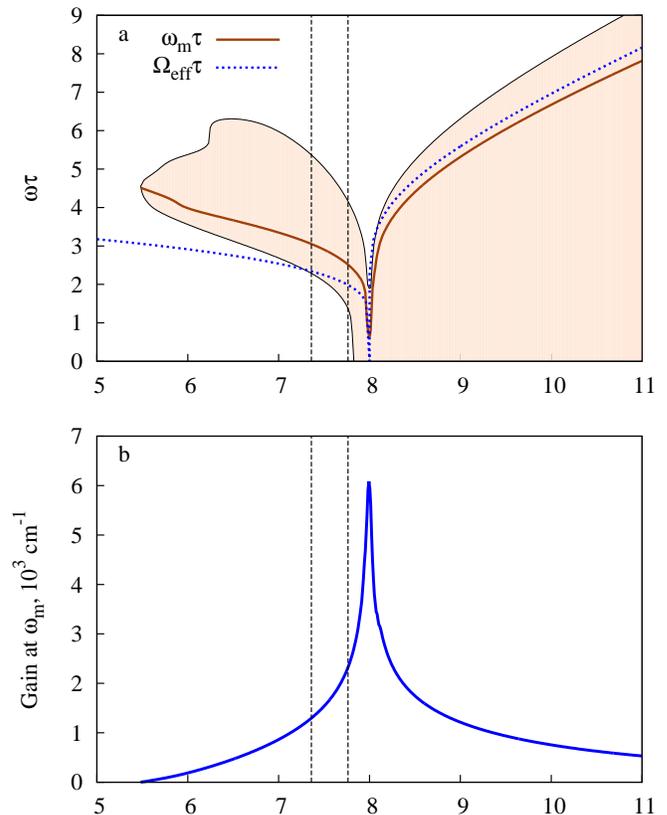}%
\caption{(a) Region of fields and frequencies corresponding to the
amplification of a weak alternating field in the superlattice
(shaded area). The solid line is the plot of frequencies
$\omega_m\tau$ at which the gain is maximal and the dashed line is
the plot of effective frequencies $\Omega_{eff}\tau$ of oscillations
of ballistic electrons. (b) Maximum possible gain versus the scaled
electric field field $\omega_B\tau$ at $\omega_c\tau=4$. The meaning
of the vertical dashed straight lines is the same as in Fig. 2.}
 \label{fig:3}
\end{figure}

Figure 3a shows the region in the parametric space ($\omega_B\tau,
\omega\tau$) that corresponds to the amplification of stimulated
radiation. As can be seen in the figure, the amplification region
is wider than the region of negative effective mass. This can
serve as an additional illustration of a statement that negative
effective mass is only one of the conditions for the appearance of
amplification \cite{Kaus59}-\cite{Kittel59}. It is obvious that
not all frequencies are amplified identically. In particular, the
maximum possible gain is reached at the frequencies $\omega_m$
shown in the same figure. As was mentioned above, amplification in
this case is directly attributed to oscillations of electrons in
the miniband. Indeed, it is seen in Fig. 3a that the dependence of
$\omega_m$ on the electric field strongly correlates with the
corresponding dependence for the effective frequency of electron
oscillations $\Omega_{eff}$.

Figure 3b shows the dependence of the gain on $\omega_B\tau$ for
the frequencies $\omega$ at which the gain is maximal (i.e., for
$\omega_m$). The maximum gain is reached near the separatrix and
at the relatively low frequencies. As was previously emphasized in
\cite{Hyart09}, the magnitude of gain is very large near the
separatrix. We now estimate the range of frequencies $\omega$.
Taking a value of about $200$ fs for the relaxation time of the
typical superlattice, we conclude that the amplified frequencies
are in the range from $500$ GHz to several terahertz. The
necessary magnetic field strength should be several tesla.

To conclude, we note that negative effective mass
of miniband electrons in superlattices can also appear due to the
application of either static electric field at conventional Bloch oscillations
\cite{Esaki70,Ignatov93} or ac pump field at parametric
amplification \cite{Hyart07} and other parametric effects
\cite{Romanov00,Shorokhov10}. However, in both indicated cases, the
mass averaged over the period of oscillations is positive and,
therefore, no population inversion exists. The application of a
perpendicular magnetic field makes electron oscillations nonlinear
and the average mass of electrons negative in the strong
nonlinearity regime. In this respect, we also would like to attract attention to the
problem of terahertz radiation amplification in the superlattice placed in a
tilted magnetic field \cite{Hyart09}. Here miniband electrons demonstrate
strongly nonlinear dynamics which is manifested in the existence of a chaotic
web \cite{Fromhold01}. Possibility of contribution of the negative effective mass
in the amplification in this situation
is a very interesting problem for further investigations.

We are thankful to N.~S.~Prudskikh for technical assistance. This
work was supported by Ministry of Education and Science of Russian
Federation (project no. 2.2665.2014), Engineering and Physical
Sciences Research Council (grant no. EP/I01490X/1), Royal Society
(UK-India Collaboration).

\end{document}